\documentclass[aps,prb,reprint,showpacs,floatfix]{revtex4-1}
\usepackage{amssymb}
\usepackage{bm}
\usepackage{graphicx}
\usepackage{natbib}
\usepackage[utf8]{inputenc}
\renewcommand\Re{\mathop{\rm Re}}
\renewcommand\Im{\mathop{\rm Im}}
\begin{document}

\title{Long-range spin-triplet proximity effect in Josephson junctions with
multilayered ferromagnets}
\author{Luka Trifunovic and Zoran Radović}
\affiliation{Department of Physics, University of Belgrade, P.O. Box 368, 11001
Belgrade, Serbia}

\begin{abstract}
We study the proximity effect in  SF'(AF)F'S and SF'(F)F'S planar junctions,
where S is a clean conventional (s-wave) superconductor, while F' and middle
layers are clean or moderately diffusive ferromagnets. Middle layers consist of
two equal ferromagnets with antiparallel (AF) or parallel (F) magnetizations
that are not collinear with magnetizations in the neighboring F' layers. We use
fully self-consistent numerical solutions of the Eilenberger equations to
calculate the superconducting pair amplitudes and the Josephson current for
arbitrary thickness of ferromagnetic layers and the angle between in-plane
magnetisations.  For moderate disorder in ferromagnets the triplet proximity
effect is practically the same for AF and F structures, like in the dirty limit.
Triplet Josephson current is dominant for $d'\approx\hbar v_F/2h'$, where $d'$
is the F' layer thickness and $h'$ is the exchange energy. Our results are in a
qualitative agreement with the recent experimental observations [T. S.  Khaire,
M. A.  Khasawneh, W. P.  Pratt, and N. O. Birge, Phys. Rev. Lett.  \textbf{104},
137002 (2010)].
\end{abstract}

\maketitle

In hybrid systems containing superconducting and ferromagnetic metals, triplet
correlations are
induced.\cite{bergeret_long-range_2001,volkov_odd_2003,bergeret_odd_2005,buzdin_proximity_2005}
In the case of homogeneous magnetization, the triplet pair amplitude has zero
total spin projection on the magnetization axis:
$F_{t0}(t-t')\propto\mid\uparrow(t)\downarrow(t')\rangle+\mid\downarrow(t')\uparrow(t)\rangle$.
For equal time $t=t'$, this function vanishes, in agreement with the Pauli
principle. Therefore, $F_{t0}$ is an odd function of the time difference or
equivalently, odd in frequency. The exchange field mixes the usual spin singlet
pairing correlations,
$F_s\propto\mid\uparrow\downarrow\rangle-\mid\downarrow\uparrow\rangle$, and the
spin triplet $F_{t0}$ because the wave functions $\mid\uparrow\downarrow\rangle$
and $\mid\downarrow\uparrow\rangle$ acquire relative phase difference in the
ferromagnet.\cite{blanter_supercurrent_2004}

Amplitudes $F_s$ and $F_{t0}$ are short-ranged and oscillate spatially in the
ferromagnet, both in the clean and the dirty limit.\cite{bergeret_odd_2005} In
the clean limit, they decay algebraically in the ferromagnet with $\hbar v_F/h$,
where $h$ is the exchange energy and $v_F$ is the Fermi velocity. In the dirty
limit, they decay exponentially with the characteristic length $\sqrt{\hbar
D/h}$, where $D=v_Fl/3$ is the diffusion coefficient, and $l$ is the electron
mean free path. However, this is not true for the clean single-channel junctions
where all pairing correlations are long-ranged and their spatial decay in the
ferromagnet is independent of the exchange field.\cite{buzdin_proximity_2005}

The situation is quite different for inhomogeneous magnetization: not only $F_s$
and $F_{t0}$ amplitudes exist but also odd triplet pair amplitude $F_{t1}$ with
$\pm1$ total spin projection on the magnetization axes emerges in the
ferromagnet.\cite{bergeret_long-range_2001,volkov_odd_2003,bergeret_odd_2005}
The triplet component $F_{t1}$ is not suppressed by exchange interaction and
penetrates into the ferromagnet over large distance on the order of $\sqrt{\hbar
D/k_BT}$ in the dirty limit ($l<\hbar v_F/h$) and moderately disordered
ferromagnets, and, likewise, over the distance $\hbar v_F/k_BT$ in the clean
limit ($l>\hbar v_F/k_BT$).

It is not difficult to understand why $F_{t1}$ triplet component is induced in
inhomogeneous ferromagnets. Consider a system where inhomogeneity is represented
by two ferromagnetic layers with noncollinear magnetizations and angle $\alpha$
between them. As we have already pointed out, in each ferromagnetic layer the
exchange field generates $F_{t0}$ from $F_s$. When $F_{t0}$ correlation
penetrates into neighboring ferromagnetic layer it mixes with $F_{t1}$ having
non-zero total spin projection due to different orientation of magnetizations.
This implies that $F_{t1}$ triplet component is generated in the presence of
inhomogeneous magnetization and is proportional to $F_{t0}\sin{\alpha}$ at the
interface between two ferromagnetic layers. Therefore, for fully developed
triplet proximity effect one of two layers should be sufficiently thin to
provide large $F_{t0}$ at the interface between the
layers.\cite{houzet_long_2007,volkov_odd_2008}

Another possibility for dominant long-range triplet component was suggested in
Refs.
\onlinecite{eschrig_theory_2003,eschrig_triplet_2008,asano_odd-frequency_2007}.
In that approach the role of thin ferromagnetic layers is replaced by
spin-active interfaces described by scattering matrix. The elements of the
scattering matrix can be considered as phenomenological parameters. Purely
microscopic approach that retains quasiparticle information at the atomic scale,
with spin-dependent scattering potentials at the interfaces, was considered in
Ref. \onlinecite{halterman_emergence_2009}.

Besides the impact on the Josephson current, another signature of odd-frequency
triplet pair correlations has been proposed recently: The density of states in
the presence of the odd-frequency pairing is enhanced, and acquires a
zero-energy
peak.\cite{yokoyama_manifestation_2007,*yokoyama_tuning_2009,braude_fully_2007}

Experimental results that may support the existence of long range triplet
amplitudes have already been
obtained.\cite{sosnin_superconducting_2006,keizer_spin_2006,anwar_supercurrents_2010}
However, in these experiments it was not possible to tune the junction
parameters, and the transition from usual singlet to long-range triplet
proximity effect has not been observed.

Quite recently, long-range triplet proximity effect has been observed in
SF'(AF)F'S junctions with F' layer thickness as a controllable
parameter.\cite{khaire_observation_2010} Here F' is a weak ferromagnetic layer
(PdNi or CuNi) and AF is synthetic antiferromagnet: an exchange-coupled trilayer
Co/Ru/Co. Previously, a similar arrangement (with homogeneous middle layer F)
has been analyzed theoretically and proposed as a good candidate for the
long-range triplet effect.\cite{houzet_long_2007,braude_fully_2007}

Junctions with F and AF middle layers were analyzed recently for the case
when F' layer thickness is much smaller than $\sqrt{\hbar
D/h'}$.\cite{volkov_odd_2010} However, within this approximation the triplet
component is not dominant and consequently results are not applicable to the
experiment (Ref. \onlinecite{khaire_observation_2010}). More interesting case
was considered by the same authors, when F' layer thickness is arbitrary but the
middle layer is homogeneous, as in Ref. \onlinecite{houzet_long_2007}. In the
dirty limit (and moderately disordered ferromagnets as we will show), results
are practically the same for F and AF structures of the middle layer. The
situation is different for the case of clean ferromagnets.

In this paper, we study the proximity effect in clean or moderately diffusive 
SF'(AF)F'S and SF'(F)F'S planar junctions, where S is a conventional (s-wave)
superconductor, while F' and middle layers are ferromagnets. Middle layer
consists of two equal ferromagnets with antiparallel (AF) or parallel (F)
magnetizations that are not collinear with magnetizations in the neighboring F'
layers.

To calculate the Josephson current and pair correlations in the clean limit and
for moderately diffusive ferromagnets, we use the Eilenberger equations
\cite{eilenberger_transformation_1968} for a junction along the $x$-axis
\begin{equation}
\label{eilenberger} \hbar v_x\partial_x \check{g}+\Big[\omega_n \hat{\tau}_3-i
\check{V}+\check{\Delta}+\hbar \langle \check{g}\rangle/2\tau, \check{g}\Big]=0,
\end{equation}
with normalization condition $\check{g}^2=\check{1}$. Disorder is characterized
by the average time $\tau=l/v_F$ between scattering on impurities.  We indicate
by $\hat{\cdots}$ and $\check{\cdots}$ $2\times2$ and $4\times4$ matrices,
respectively. Here, $v_x=v_F\cos\theta$ where $\theta$ is an angle between the
Fermi velocity and the $x$-axis, $\hat\tau_i$ are the Pauli matrices in the
particle-hole space, the brackets $\langle\ldots\rangle$ denote angular
averaging over the Fermi surface (integration over $\theta$), and
$[\;,\;]$ denotes a commutator. The quasiclassical Green functions are
given by
\begin{equation}
  \check{g}=\left(
  \begin{array}{cccc}
    g_s+\bm{g_t\cdot}\hat{\bm\sigma} & (f_s+\bm{f_t\cdot}\hat{\bm\sigma})i\hat{\sigma}_y \\
    -(\tilde{f_s}+\tilde{\bm{f_t}}\cdot\hat{\bm\sigma}^*)i\hat{\sigma}_y &
	-(g_s+\bm{g_t}\cdot\bm{\hat{\sigma}^*})
  \end{array} \right),
  \label{gmatrix}
\end{equation}
where $\hat{\bm\sigma}=(\hat{\sigma}_x,\hat{\sigma}_y,\hat{\sigma}_z)$ are the
Pauli matrices in spin space. With this parametrization,
\cite{eschrig_triplet_2008} it is clear that $g_s$ and $f_s$ remain unchanged
under rotation of magnetizations, while $\bm{g_t}$ and $\bm{f_t}$ transform like
ordinary vectors. The conjugate Green's functions satisfy $\tilde
f_s(\omega_n)=f_s^*(-\omega_n)$ and $\tilde{\bm{f_t}}
(\omega_n)=-\bm{f_t}^*(-\omega_n)$.\cite{eschrig_symmetries_2007} The exchange
field term in ferromagnets is given by $\check{V}=\Re[
\bm{h}(x)\cdot\hat{\bm{\sigma}}]+i\hat{\tau}_3 \Im[
\bm{h}(x)\cdot\hat{\bm{\sigma}}]$ and the pair potential in superconductors is
$\check\Delta=(\hat\tau_{+}\Delta+\hat\tau_{-}\Delta^*)\hat{\sigma}_y$, where
$\hat\tau_{\pm}=\hat\tau_x\pm i\hat\tau_y$. The exchange field $\bm h$ has
the following $x$-dependence
\begin{equation}
  \bm h=\left\{
  \begin{array}{ll}
    h'(-\sin\alpha\;\mathbf{y}+\cos\alpha\;\mathbf{z}), & -d-d'<x<-d, \\
    h\;\mathbf{z}, & -d\le x<0, \\
    \pm h\;\mathbf{z}, & 0\le x<d, \\
    h'(\sin\alpha\;\mathbf{y}+\cos\alpha\;\mathbf{z}), & d\le x<d+d', \\
  \end{array}\right.
  \label{exchangefield}
\end{equation}
where $d'$ and $2d$ are F' and (F) or (AF) thickness, respectively. The angle
between magnetizations in F' and neighboring ferromagnetic layers is $\alpha$;
the plus (minus) sign is for F (AF) middle layer. In the absence of out-of-plane
magnetization, the amplitude $(\bm{f_t})_x=0$.

\begin{figure}
\includegraphics[width=8.5cm]{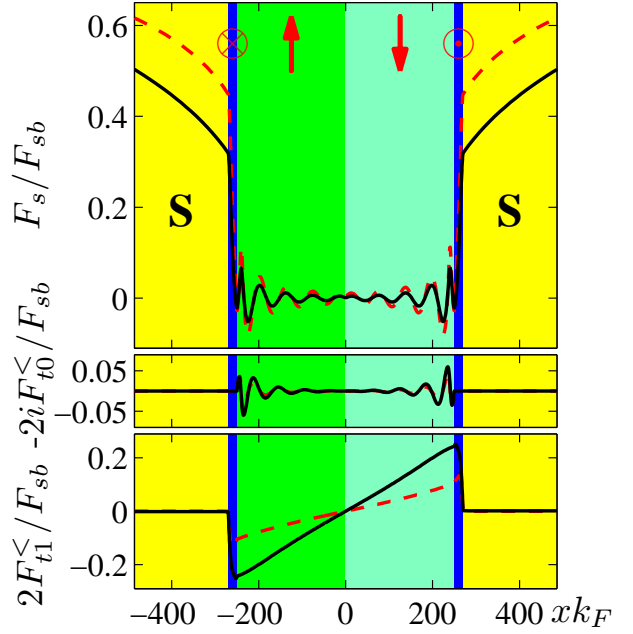}
\caption{(color online) Spatial dependence of singlet and triplet pair
amplitudes $F_s$, $F_{t0}^<$ and $F_{t1}^<$, normalized to the bulk singlet
amplitude $F_{sb}$, for $T/T_c=0.1$, $h'/E_F=0.05$, $h/E_{F}=0.1$,
$2d=500k_F^{-1}$, $d'=20k_F^{-1}$, $\alpha=\pi/2$ and two values of the mean
free path: $l=\infty$ (dashed curves) and $l=200k_F^{-1}$ (solid curves). All
amplitudes are calculated for the ground state, $\phi=0$. The SF'(AF)F'S
junction geometry is shown in the background; arrows and circles show
orientation of magnetizations in each layer.}
\label{pairamplitudes}
\end{figure}

The supercurrent flowing through the junction is given by the normal Green
function
\begin{equation} \label{struja} I(\phi)=\pi e
  N(0)Sk_BT\Im\sum_{\omega_n}\langle v_x g_s(v_x)\rangle ,
\end{equation}
where $\phi$ is the macroscopic phase difference across the junction, $N(0)$
is the density of states per spin at the Fermi surface, and S is the area of the
junction. Equal-time pair amplitudes are defined in terms of anomalous Green
functions as
\begin{eqnarray}
F_s&=&-i\pi N(0)
k_BT\sum_{\omega_n}\langle f_s\rangle,\\
F^{<}_{t0}&=& \pi N(0)
k_BT\sum_{\omega_n<0}\langle(\bm{f_t})_z\rangle,\\
F_{t1}^{<}&=& -i\pi N(0)
k_BT\sum_{\omega_n<0}\langle(\bm{f_t})_y\rangle.
\end{eqnarray}
Equal-time triplet amplitudes identically vanish according to the Pauli
principle, hence we defined auxiliary functions using summation over negative
frequencies only. The spatial variation of the time-dependent triplet pair amplitudes
is qualitatively the same as for auxiliary functions.

We consider the case of fully transparent interfaces, i.e., strong proximity
effect. The opposite case of low transparency was considered in Ref.
\onlinecite{volkov_odd_2010}. Using continuity of the Green functions at
interfaces, Eqs. (\ref{eilenberger}) are solved iteratively with the
self-consistency condition $\Delta=\lambda F_s$. Iterations are performed until
self-consistency is reached, starting from the stepwise approximation for the
pair potential
$\Delta=\Delta(T)[e^{-i\phi/2}\Theta(-x-d-d')+e^{i\phi/2}\Theta(x-d-d')]$. The
temperature dependence of the bulk pair potential is given by $\Delta (T)=\Delta
(0)\tanh \left( 1.74\sqrt{T_{c}/T-1}\right)$. For arbitrary mean free path in
ferromagnets we employ the iterative procedure starting from the clean limit.

Figure~\ref{pairamplitudes} shows the spatial variation of the pair amplitudes for
the SF'(AF)F'S junction geometry with magnetizations in F' layers orthogonal to
the neighboring middle layers, $\alpha=\pi/2$. In this case the 0-state is the
ground state. Transition to the $\pi$-state can be tuned with relative
orientation of magnetizations in F' layers ($\pi$-state is the ground state for
parallel magnetizations in the two F' layers). For corresponding SF'(F)F'S
junctions pair amplitudes are practically the same. For $\phi=0$, the singlet
and the long-range triplet amplitudes, $F_s$ and $F_{t1}$, are real while the
short-range triplet amplitude $F_{t0}$ is imaginary.

The F' layer thickness $d'=20k_F^{-1}$ is chosen to give the maximum triplet
current for moderate disorder in ferromagnets ($l=200k_F^{-1}$). This explains
the large difference between amplitudes of the long-range triplet component in
the clean and moderately disordered case, Fig.~\ref{pairamplitudes}.

We observe oscillatory decay of $F_s$ and $F_{t0}^<$ amplitudes, dependent on
the exchange field. In contrast, long range triplet component, $F_{t1}^<$, is
monotonic in the middle layer and its decay length is independent of the
exchange field magnitude. Thin F' layers are considered as weak ferromagnets,
$h'/E_F=0.05$, and thick middle layers, $2d=500k_F^{-1}$, represent strong
ferromagnets. For illustration of pair amplitudes, $h/E_F=0.1$ is used in order
to have larger period of spatial oscillations, although $h/E_F=0.3$ is used in
other illustrations. All amplitudes are normalized to the bulk singlet pair
amplitude $F_{sb}=2\pi
N(0)k_BT\sum_{\omega_n}\Delta/\sqrt{\omega_n^2+\Delta^2}$.

\begin{figure}
\includegraphics[width=8.5cm]{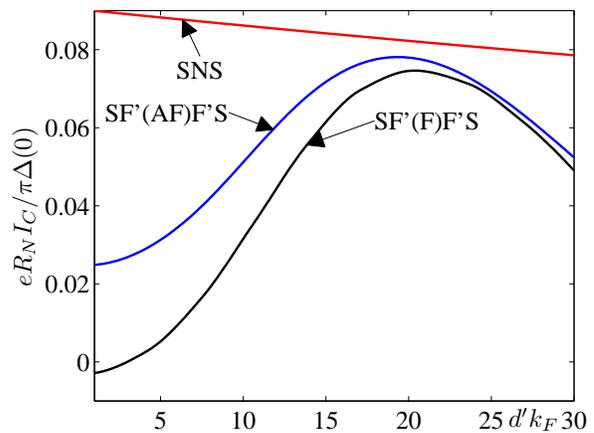}
\caption{(color online) Dependence of the Josephson critical current on the
thickness $d'$ of F' ferromagnetic layer , for $2d=500k_F^{-1}$, $T/T_c=0.1$,
$h'/E_F=0.05$, $h/E_{F}=0.3$, $l=200k_F^{-1}$, and for three types of junctions:
SNS, SF'(AF)F'S, and SF'(F)F'S.}
\label{Ic(d')}
\end{figure}

Next we examine the Josephson critical current dependence on F' layer thickness.
This quantity is actually measured in the
experiment.\cite{khaire_observation_2010} Figure~\ref{Ic(d')} illustrates
$I_C(d')$, normalized to the resistance $R_N=2\pi^2\hbar/Se^2k_F^2$, for three
types of junctions: SF'(AF)F'S, SF'(F)F'S and SNS, where N is the corresponding
normal nonmagnetic metal ($h=h'=0$). Here, we take mean free path
$l=200k_F^{-1}$ in ferromagnetic and N metals.

In the clean limit, $I_C(d')$ curves for SF'(AF)F'S and the corresponding SNS
junctions coincide,\cite{blanter_supercurrent_2004} while in the dirty limit
$I_C(d')$ curves for SF'(AF)F'S and SF'(F)F'S curves are practically the
same.\cite{volkov_odd_2010} For intermediate disorder in ferromagnets, the
critical current is always larger in SF'(AF)F'S than SF'(F)F'S junctions,
Fig.~\ref{Ic(d')}.

We find for moderate disorder in ferromagnets that the largest critical current
is almost as big as for the corresponding SNS junction. Note that in the dirty
limit maximum triplet critical current is only $12\%$ of the critical current of
the corresponding SNS junction.\cite{volkov_odd_2010} The position of maxima of
$I_C(d')$ strongly depends on $h'$, $d'_{max}\approx\hbar v_F/2h'$. It is
practically independent of $h$ and $d$, in agreement with the results of Ref.
\onlinecite{volkov_odd_2010}.

\begin{figure}
\includegraphics[width=8.5cm]{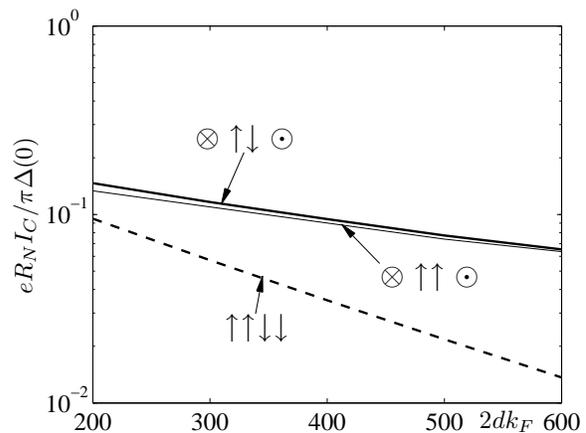}
\caption{Dependence of the Josephson critical current on the middle layer
thickness $2d$, for $d'=20k_F^{-1}$, $T/T_c=0.1$, $h'/E_F=0.05$, $h/E_{F}=0.3$,
$l=200k_F^{-1}$, and for two values of the misorientation angle: $\alpha=\pi/2$
(thick and thin solid lines for AF and F structures) and for $\alpha=0$ (AF
structure, dashed line). Arrows and circles show orientation of magnetizations
in ferromagnetic layers.}
\label{Ic(d)}
\end{figure}

The critical Josephson current dependence on the middle layer thickness is shown
in Fig.~\ref{Ic(d)}. The F' layer thickness is set to $d'_{max}=20k_F^{-1}$. We
consider $I_C$ dependence on $2d$ for two values of misorientation angle:
$\alpha=\pi/2$ (fully developed triplet proximity effect, solid lines) and
$\alpha=0$ (no long range triplet component, dashed line). For thick middle
layer, $2d=600k_F^{-1}$, in both AF and F geometries, $I_C$ is an order of
magnitude larger when the triplet component $F_{t1}$ is present. These results,
Figs.~\ref{Ic(d')} and \ref{Ic(d)}, are in a qualitative agreement with
experimental observation.\cite{khaire_observation_2010}

\begin{figure}
\includegraphics[width=8cm]{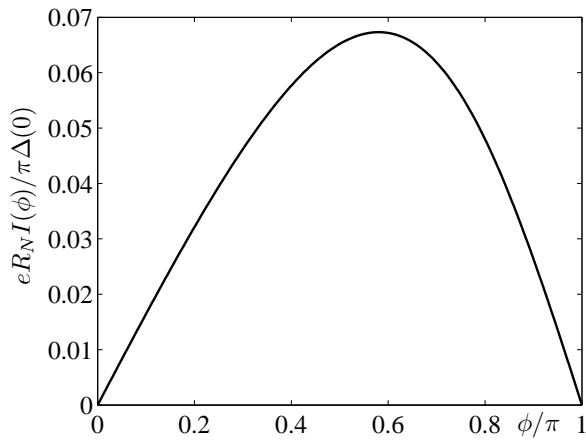}
\caption{The current-phase relation $I(\phi)$ for SF'(AF)F'S junction: 
$d'=20k_F^{-1}$, $2d=600k_F^{-1}$, $T/T_c=0.1$, $h'/E_F=0.05$,
$h/E_{F}=0.3$, $l=200k_F^{-1}$, and $\alpha=\pi/2$.}
\label{I(phi)}
\end{figure}

The current-phase relation is almost harmonic for a moderately disordered
SF'(AF)F'S junction ($l=200k_F^{-1}$), Fig.~\ref{I(phi)}. Here, the F' layer
thickness is optimal for long-range triplet Josephson effect; i.e., the usual
singlet Josephson critical current is an order of magnitude smaller. For the
in-plane magnetizations considered here we did not find any unusual $I(\phi)$
dependence, in agreement with Refs.
\onlinecite{houzet_long_2007,volkov_odd_2010}.

However, we expect unusual $I(\phi)$ dependence for the samples used in the
experiment (Ref. \onlinecite{khaire_observation_2010}). Supercurrent could be
observed even for $\phi=0$, if one takes into account the possibility of
inhomogeneous out-of-plane magnetisations in PdNi layers and the fact that
transport properties of the minority and majority electrons at the Fermi surface
in Co layers are very different. Quasi-classical approximation we used in this
paper does not allow for the latter possibility; this is the principal
limitation of our approach.

A new long-range triplet component,
$(\bm{f_t})_x=(f_{\uparrow\uparrow}-f_{\downarrow\downarrow})/2$, is generated
in the case of non-zero relative longitude angle $\chi$ between magnetizations
in two F' layers (i.e., inhomogeneous out-of-plane magnetization). This can be
readily seen from Eq. (\ref{eilenberger}) for $(\bm{f_t})_x$ component. The
presence of $(\bm{f_t})_x$ implies different triplet Josephson current flow for
majority ($I_\uparrow$) and minority ($I_\downarrow$) electrons. In the circuit
theory approximation,\cite{braude_fully_2007} it was obtained that
$I_{\uparrow,\downarrow}=C_{\uparrow,\downarrow}\sin{(\phi\pm \chi)}$, where
$C_{\uparrow,\downarrow}$ are proportional to densities of states and Fermi
velocities of spin subbands. Although the case of very different Fermi
velocities of the subbands is not accessible within the circuit theory
approximation, it is reasonable to assume that similar current-phase relations
are valid.  Consequently, there is a finite supercurrent at zero phase
difference, as was predicted for the half-metallic middle
layer.\cite{braude_fully_2007}

It would be very interesting to measure the $I(\phi)$ dependence for the
samples used in the experiment (Ref. \onlinecite{khaire_observation_2010}),
since the out-of-plane magnetizations in thin PdNi layers are their typical
feature.\cite{petkovi_direct_2009} A non-zero supercurrent for $\phi=0$ could
provide an unambiguous evidence of the triplet proximity effect.

We acknowledge useful discussions with Norman Birge, Marco Aprili, Ivana
Petković, Mihajlo Vanević and Ivan Božović. The work was supported by the
Serbian Ministry of Science, Project No. 141014. 
\bigskip
\bibliography{reference}
\bigskip
\end{document}